\documentclass[12pt]{article}
\usepackage{amsmath}
\usepackage{amssymb}
\usepackage{float}
\usepackage{graphicx}
\usepackage{tabularx,ragged2e,booktabs,caption}
\usepackage{caption}
\usepackage{subcaption}
\usepackage{hyperref}
\usepackage{slashed}
\usepackage{multicol}
\usepackage[margin=1.0in,footskip=0.25in]{geometry}

\begin{document}
\begin{center}
{\bf Swampland Criteria in Slotheon Field Dark Energy}
\end{center}
\begin{center}
 {\bf Upala Mukhopadhyay $^\dagger$, Debasish Majumdar $^\ddagger$}
\end{center}
\vskip 0.5mm
\begin{center}
Astroparticle Physics and Cosmology Division  \\
Saha Institute of Nuclear Physics, HBNI  \\
1/AF Bidhannagar, Kolkata 700064, India  \\
\end{center}
\bigskip
\bigskip
\bigskip
\bigskip
\begin{center}
{\bf Abstract} 
\end{center}
{\small We explore in this work whether the Slotheon model of Dark Energy obeys the Swampland criteria of string theory. Since de Sitter vacuum is very difficult to construct in string theory the cosmological constant as an explanation of Dark Energy is almost ruled out in string theory as it involves a scalar potential $V$ with positive local minimum that ends up to a stable (or meta stable) de Sitter (ds) vacuum. In quintessence model however if the derivative of the scalar potential $V$ ($\nabla V$) is small and $\frac{|\nabla V|}{V} \sim {\cal {O}}(1)$  then in this situation the potential $V$ can be positive but the scalar field may not be at the minimum. For a consistent quantum theory of gravity the theory should not have any ds or meta stable ds vacua. In this regard the Swampland criterion is proposed which any low energy theory should obey to be consistent with quantum theory of gravity. This criterion is written as $|\nabla V|/V > c \sim {\cal {O}}(1)$. In this work we consider a scalar field model for Dark Energy namely the Slotheon Dark Energy model inspired by the theories of extra dimensions and show that this Dark Energy model agrees better with the Swampland criteria than the quintessence Dark Energy model. }
\vskip 7cm
\noindent  email: $^\dagger$ upala.mukhopadhyay@saha.ac.in \\
$~~~~~~~~~$ $^\ddagger$ debasish.majumdar@saha.ac.in
\newpage

\section{Introduction}

The scalar field theories such as quintessence as also the scalar fields inspired by the theories of extra dimensions are considered in the literature to account for the late time acceleration of the Universe and the Dark Energy of the Universe which is thought to have caused this acceleration. The basis of such theories for Dark Energy is generally the Einstein's theory of general relativity which appears to work well, below the Planck scale. But beyond the Planck scale it is debatable whether such theories can connect to the more robust quantum theory of gravity in a string theory landscape. It appears that there exists an even bigger string Swampland \cite{swampland} where some effective field theories coupled to gravity are inconsistent with the quantum theory of gravity. This has arisen from the difficulties in string theory in constructing the de Sitter vacuum and to find the possibility of the absence of de Sitter like vacuum in a consistent quantum theory of gravity. Therefore it may be worthwhile to investigate whether a low energy theory of gravity obeys the string Swampland criteria such that this theory can be attributed to be a low energy theory of a consistent quantum theory of gravity and thus can be embedded in string theory. Therefore the string theory criteria can be used to
constrain the Dark Energy models that originate from a scalar field theory. There are works in literature that address Swampland criteria and quintessence models [\cite{5} - \cite{coupling}].

In this work we explore whether the Slotheon model \cite{upala,debabratada,germani} for Dark Energy obeys the 
string Swampland criteria. For a detail discussion of the Slotheon field Dark Energy model one may see Ref. \cite{upala,debabratada}. The string Swampland criteria for an effective field theory to be consistent with the string theory, are given by
\begin{itemize}
\item The range $d$ traversed by the scalar fields should obey the bound
$|\Delta \pi| < d \sim {\cal {O}}(1)$ (in reduced Planck units) \cite{criteria1}.
\item The quantity $|\nabla_\pi V|/V > c \sim {\cal {O}}(1)$ in reduced 
Planck units \cite{criteria2}, where $V$ is the potential of the scalar field 
$\pi$ and $V>0$. This means that the derivative of the potential $V$ of the scalar field 
has a finite minima (a lower bound). 
\end{itemize}

A refined form of the second Swampland criterion can also be written as \cite{connection}
\begin{itemize}
\item The potential $V(\pi)$ of the scalar field $\pi$ must satisfy either  $|\nabla_\pi V|/V > c \sim {\cal {O}}(1)$ or $\nabla^2_{\pi}V/V<-c' \sim {\cal {O}}(1)$ in reduced Planck units. 
\end{itemize}

The second criterion is from the fact that it is difficult to construct the 
de Sitter vacuum. It is argued in \cite{l.heisenberg} that the second criterion is relevant for Dark 
Energy. Therefore in the present context the second condition is important.
In this work we investigate the consistency of  the Dark Energy from 
Slotheon scalar field theory with this string Swampland criterion. In this regard
we explore the variations of Dark Energy equation of state parameters. We 
consider a generalised thawing model \cite{thawing} of Dark Energy \cite{parametrisation} to construct the variations of Dark Energy equation of state where the 
experimental bounds are considered by choosing the parameter $\omega_0$ from 
Euclid \cite{euclid} simulated data. We then compute the variations of $w(z)$ 
with $z$ for Slotheon Dark Energy scalar field model. This is done for different chosen values of $\lambda=\frac{M_{\rm pl} V'}{V}$, where $M_{\rm pl}$ denotes the reduced Planck mass and $V$ is Slotheon scalar field potential. These are then compared with the thawing Dark Energy limits to test the Swampland criteria for Slotheon Dark Energy model. We repeat this comparison for quintessence scalar field model and found that Slotheon Dark Energy model satisfies Swampland criteria better than the quintessence model.

In an earlier work L. Heisenberg et al \cite{l.heisenberg} have performed a similar test for Swampland criterion for scalar quintessence model. In that work they have considered a quintessence field and discussed about the Swampland criteria for the quintessence field Dark Energy. In doing so they have taken the experimental bounds by writing the variations of Dark Energy equation of state in the usual CPL \cite{CPL} parametrisation form and then translate it to obtain an upper bound of a reconstructed equation of state $\omega(z)$ ($\omega(z) \sim \omega_0 + \frac{z}{1+z} \omega_a$). Here there are two parameters namely $\omega_0$ and $\omega_a$ \footnote{CPL parametrisation finds its importance in varied contexts such as constraining the sum of neutrino masses in dynamical Dark Energy models \cite{3}}. For this purpose L. Heisenberg et al have taken the constraints on SNeIa, CMB, BAO and $H_0$ measurements data (Fig. 21 of Ref. \cite{9of_DE}). They have also repeated their analyses for Swampland criteria for comparison with Euclid simulation of future data.

We have performed in this work a similar analysis of Swampland conditions in case of Slotheon field Dark Energy and standard quintessence field Dark Energy using Euclid simulated data but in our case we parametrise our Dark Energy equation of state with a generalised thawing Dark Energy model \cite{parametrisation}. We also repeat our analyses with the constraints used by L. Heisenberg e al \cite{l.heisenberg}.

We also mention here that in a recent work \cite{brahma} Brahma et al has done a similar study with the cubic Galileon term (($\nabla \pi)^2 \square \pi$) in their chosen Galileon action for Dark Energy. For the experimental bounds on Dark Energy equation of state they also consider (similar to that in Ref. \cite{l.heisenberg}) the CPL parametrisation. But in our work we explore the Swampland criterion for Slotheon field Dark Energy model. Also as mentioned earlier, for the experimental bounds we consider the Dark Energy equation of state in a generalised thawing model. Although both Galileon and Slotheon field arise from the DGP model (Dvali, Gabadadze, Porrati model \cite{DGP}) in its decoupling limit $r_c\longrightarrow\infty$ \cite{dclt1,dclt2}($r_c=\frac{M_{\rm pl}^2}{2 M^3_5}$, $M_{\rm pl}$ and $M_5$ are bulk and brane Plank masses respectively; $r_c$ separates the 4-D and 5-D regimes), the Galileon field is described by a scalar field $\pi$ from the DGP theory in Minkowski space time that obeys the shift symmetry $\pi \rightarrow \pi + a +b_\mu x^\mu$. The Slotheon field on the other hand arises when the Galileon transformation is generalised to curved space time and obeys the curved Galileon transformation \cite{germani}
\begin{equation}
\pi(x) \rightarrow \pi(x) +c + c_a \int_{\gamma,x_0}^x \xi^a\,\,, 
\end{equation}
where $\xi^a$ is set of Killing vectors and $x_0$ is a reference point connected to $x$ by a curve $\gamma$ while $c$ and $c_a$ are a constant and a  constant vector respectively. In fact the slow rolling criteria are more favoured in Slotheon field model than the quintessence scalar field model, since the former induces an extra friction that enables the slow rolling nature to be naturally realised \cite{upala}. We also mention here in passing that L. Heisenberg et al \cite{horndeski} has performed an analysis considering a form of scalar-tensor theory with a Horndeski Lagrangian that includes cubic and quartic interactions of the scalar field. But their analysis does not include the present work with Slotheon field.

This paper is organised as follows. In Section 2, we furnish the action for quintessence and Slotheon field and provide the necessary mathematical equations as also the dimensionless variables for both the cases that are required to calculate Dark Energy equation of state parameters of both the fields. Section 3 gives a brief account of the generalised thawing model as discussed in Ref. \cite{parametrisation} for choosing the Dark Energy equation of state parametrisation used in this work to obtain the variations of $\omega(z)$ with redshift $z$ considering the experimental constraints. In Section 4 we furnish our calculations and results. We compute the variations of Dark Energy equation of state in the framework of Slotheon model for different values of $\lambda=\frac{M_{\rm pl} V'}{V}$ and compare them with those obtained using experimental constraints from Euclid simulated data \cite{euclid} and generalised thawing Dark Energy parametrisation. We also repeat the process for standard quintessence field of Dark Energy. Finally in Section 5 we give a summary and discussion.
\section{Quintessence and Slotheon Fields}

In this section we consider both the standard quintessence field $\phi$ and Slotheon field $\pi$ and calculate the equation of state of these fields.

\noindent {\underline{\it Quintessence Field}}

The action of quintessence field is given as \cite{quint tsuji}
\begin{equation}
S=\int d^4 x \sqrt{-g} \left[\frac{1}{2} M_{\rm pl}^2 R -\frac{1}{2} g^{\mu\nu}  \phi_{;\mu} \phi_{;\nu} -V(\phi) \right] +S_m\,\,, \label{action quint}
\end{equation}
where $M_{\rm pl}$ is the reduced Planck mass, $g_{\mu\nu}$ is the metric while $g$ is the determinant of the metric and $R$ is Ricci scalar. In the above, $S_m$ is the action of standard matter field, $V(\phi)$ is the potential for the quintessence field $\phi$ and $\phi_{;\mu}$ denotes the covariant derivative of $\phi$. 

By varying the action given in Eq. (\ref{action quint}) with respect to the metric and $\phi$ respectively we obtain,
\begin{eqnarray}
3 M_{{\rm pl}}^2 H^2 &=& \rho_m  + \dfrac{\dot{\phi}^2}{2} + V(\phi)\,\,, \label{EE1 quint}\\
M_{{\rm pl}}^2 (2\dot{H} + 3H^2) &=& -\dfrac{\dot{\phi}^2}{2} +V(\phi)\,\,, \label{EE2 quint} \\
\ddot{\phi} + 3H\dot{\phi} + V_{\phi}&=&0\,\,. \label{EE3 quint}
\end{eqnarray}
In the above $\dot{A}$, $\ddot{A}$ denote time derivative of $A$ and double time derivative of $A$ respectively. Derivative of potential $V(\phi)$ w.r.t. $\phi$ is given as $V_\phi$ while $\rho_m$ denotes the matter energy density. 
In order to obtain the dynamics of the system, it is convenient to introduce the following dimensionless variables,
\begin{eqnarray}
x & = & \dfrac{\dot{\phi}}{\sqrt{6}H M_{{\rm pl}}}\,\,, \label{x quint}\\
y & = & \dfrac{\sqrt{V(\phi)}}{\sqrt{3} H M_{{\rm pl}}}\,\,,\label{y quint}\\
\lambda & = & -M_{{\rm pl}} \dfrac{V_\phi}{V(\phi)}\label{l quint}\,\,.
\end{eqnarray}

With these, Eqs.(\ref{EE1 quint})-(\ref{EE3 quint}) can be written as the following autonomous set of equations \cite{sami},
\begin{eqnarray}
\dfrac{dx}{dN} &=& -3 x + \frac{\sqrt{6}}{2} \lambda y^2+\frac{3}{2} x \left[ (1-\omega_m)x^2+(1+\omega_m)(1-y^2)\right]\,\,,\label{auto1 quint}\\
\dfrac{dy}{dN} & = &  -\sqrt{\frac{3}{2}} \lambda x y + \frac{3}{2}y\left[(1-\omega_m)x^2+(1+\omega_m)(1-y^2)\right]\,\,,\\
\dfrac{d\lambda}{dN} & = & -\sqrt{6} x \lambda^2 \left(\frac{V V_{\phi\phi}}{V_\phi^2}-1 \right)\,\,.\label{auto3 quint}
\end{eqnarray}
Here $V_{\phi\phi}$ is the double derivative of $V(\phi)$ w.r.t. $\phi$, $\omega_m$ represents the equation of state for the matter field and $N={\rm ln}a$ ($a$ being the scale factor of the Universe) is number of e-foldings. 

Effective equation of state parameter $\omega_{\rm eff}$ and equation of state parameter of Dark Energy $\omega_\phi$ for this system are obtained from Einstein's equations (Eqs. (\ref{EE1 quint} - \ref{EE3 quint})) and are given as
\begin{eqnarray}
\omega_{\rm eff} &=& \frac{p_{\rm total}}{\rho_{\rm total}} = \frac{p_{m} +p_\phi}{\rho_m +\rho_\phi}=-1-\frac{2 \dot{H}}{3 H^2}\,\,, \label{weff quint}\\
\omega_\phi &=& \frac{\omega_{\rm eff}}{\Omega_\phi}\,\,.\label{w quint}
\end{eqnarray}
In the above, density parameter $\Omega_\phi$ of the field $\phi$ is defined as $\Omega_\phi=\frac{\rho_\phi}{\rho_c}$ while $\rho_c$ is the critical density of the Universe and $\rho_\phi$ is the energy density of the quintessence field. Needless to mention that we obtain $\omega_{eff}$ and $\omega_\phi$ in terms of the dimensionless variables (Eqs. (\ref{x quint} - \ref{l quint})). 

\noindent {\underline{\it Slotheon Field}}

Slotheon field model is a scalar field model inspired by the theories of extra dimensions, which is a class of modified gravity models. Slotheon field model is followed from  Dvali, Gabadadze and Porrati (DGP) model with one extra dimension.

Action of the Slotheon field $\pi$ is given as \cite{debabratada}
\begin{equation}
S = \int d^4x \sqrt{-g} \left[\dfrac{1}{2} \left(M_{{\rm pl}}^2 R - \left(g^{\mu\nu} - \dfrac{ G^{\mu\nu}}{M^2} \right)\pi_{;\mu} \pi_{;\nu}\right) - V({\pi})\right)+S_m \,\,. \ \label{action sloth}
\end{equation}
It can be noted from the above action that without the term $\frac{G^{\mu\nu}}{2M^2} \pi_{;\mu}\pi_{;\nu}$ in Eq. (\ref{action sloth}), both the actions of Eqs. (\ref{action quint}) and (\ref{action sloth}) are identical. In the Slotheon action (Eq. (\ref{action sloth})) $M$ represents an energy scale, $V(\pi)$ is the potential for Slotheon scalar field $\pi$, $G^{\mu\nu}$ denotes the Einstein's tensor and all the other notations are same as,  those in the standard quintessence scalar field case. 

Einstein's equations and equation of motion of Slotheon field $\pi$ are obtained by varying the action of Eq. (\ref{action sloth}) with respect to the metric $g^{\mu\nu}$ and $\pi$ respectively and are given as follows
\begin{eqnarray}
3 M_{{\rm pl}}^2 H^2 &=& \rho_m  + \dfrac{\dot{\pi}^2}{2} + \dfrac{9 H^2 \dot{\pi}^2}{2M^2} + V(\pi)\,\,, \label{EE1 sloth}\\
M_{{\rm pl}}^2 (2\dot{H} + 3H^2) &=& -\dfrac{\dot{\pi}^2}{2} +V(\pi) + (2\dot{H} + 3H^2) \dfrac{ \dot{\pi}^2}{2M^2} + \dfrac{2 H\dot{\pi}\ddot{\pi}}{M^2}\,\,, \label{EE2 sloth}\\
0 &=& \ddot{\pi} + 3H\dot{\pi} + \dfrac{3 H^2}{M^2}\left(\ddot{\pi} + 3H\dot{\pi} + \dfrac{2\dot{H}\dot{\pi}}{H}\right) + V_{\pi}\,\,. \label{EE3 sloth}
\end{eqnarray}
Here $V_\pi$ is the derivative of potential $V(\pi)$ w.r.t. $\pi$ and all the other symbols are same as in Eqs. (\ref{EE1 quint} - \ref{EE3 quint}). 

Similar to the standard quintessence case, here too it is convenient to introduce some dimensionless variables to study the evolution of the system and the variables are defined as
\begin{eqnarray}
x & = & \dfrac{\dot{\pi}}{\sqrt{6}H M_{{\rm pl}}}\,\,, \label{x}\\
y & = & \dfrac{\sqrt{V(\pi)}}{\sqrt{3} H M_{{\rm pl}}}\,\,,\label{y}\\
\lambda & = & -M_{{\rm pl}} \dfrac{V_\pi}{V(\pi)}\,\,,\label{l}\\
\epsilon & = & \dfrac{H^2}{2M^2}\,\,. \label{e}
\end{eqnarray}

\noindent Using these dimensionless variables (Eqs. (\ref{x} - \ref{e})) in the Eqs. (\ref{EE1 sloth} - \ref{EE3 sloth}), the following autonomous system of equations are constructed,
\begin{eqnarray}
\dfrac{dx}{dN} &=& \frac{P}{\sqrt{6}}-x \frac{\dot{H}}{H^2}\,\,,\label{auto1}\\
\dfrac{dy}{dN} & = & -y \left(\sqrt{\dfrac{3}{2}} \lambda x + \dfrac{\dot{H}}{H^2}\right)\,\,,\\
\dfrac{d\lambda}{dN} & = & -\sqrt{6} x \lambda^2 \left(\frac{V V_{\pi\pi}}{V_\pi^2}-1 \right)\,\,,\\
\dfrac{d\epsilon}{dN} & = & 2 \epsilon \dfrac{\dot{H}}{H^2}\,\,.\label{auto last}
\end{eqnarray}
In the above, $V_{\pi\pi}$ denotes the double derivative of potential $V(\pi)$ w.r.t. $\pi$ and 
\begin{eqnarray}
P &=&\frac{3 (12 \sqrt{6} x^3 \epsilon + 
   y^2 \lambda + 
   \sqrt{6} x (-1 -6 \epsilon y^2))}{1 + 6 \epsilon (1 + x^2 (-1 + 18 \epsilon))} \nonumber \\
   & & +\frac{-18x^2 y^2 \epsilon \lambda}{1 + 6 \epsilon (1 + x^2 (-1 + 18 \epsilon))}\,\,,
\end{eqnarray}
\begin{eqnarray}      
\frac{\dot{H}}{H^2} &=& \frac{- 
 3 x^2 (1 + 6 \epsilon) (1 + 
    18 \epsilon) + (1 + 6 \epsilon) (-3 + 
    3 y^2)}{2+12\epsilon(1+x^2(-1+18\epsilon))}  \nonumber\\
    & & + 
 \frac{12 \sqrt{6}
   x \epsilon y^2 \lambda}{2 + 12 \epsilon (1 + x^2 (-1 + 18  \epsilon))}\,\,.
\end{eqnarray}

The effective equation of state parameter for Slotheon field is obtained from Eqs. (\ref{EE1 sloth}) and (\ref{EE2 sloth}) and is given as
\begin{equation}
\omega_{\rm eff} = -1-\frac{2 \dot{H}}{3 H^2}\,\,,
\end{equation}
which is similar to Eq. (\ref{weff quint}). Also the equation of state of Dark Energy $\omega_\pi$ for Slotheon field $\pi$ is defined as
\begin{equation}
\omega_\pi=\frac{\omega_{\rm eff}}{\Omega_\pi}\,\,,
\end{equation}
where $\Omega_\pi$ is the density parameter of the Slotheon field Dark Energy.

We thus obtain the Dark Energy equation of state parameters for standard quintessence model of Dark Energy ($\omega_\phi$) and Slotheon field Dark Energy model ($\omega_\pi$) in terms of the dimensionless variables defined in Eqs. (\ref{x quint} - \ref{l quint}) and Eqs. (\ref{x} - \ref{e}) respectively. The evolutions of $\omega_\phi$ and $\omega_\pi$ with redshift $z$ can now be calculated by solving Eqs. (\ref{auto1 quint} - \ref{auto3 quint}) and Eqs. (\ref{auto1} - \ref{auto last}) respectively with proper initial conditions. We choose the initial conditions at early matter dominated epoch i.e., at redshift $z\simeq 1100$ and consider thawing Dark Energy models for both the Slotheon and
quintessence field. In a thawing Dark Energy model, equation of state of Dark Energy has a frozen initial value $-1$ and then deviates from the frozen value with time. The thawing models of Dark Energy demand the initial value of $x$ (Eq. (\ref{x quint}), (\ref{x})) to be close to zero. Hence we choose a very small initial value of $x$. The initial value of $y$ (Eq. (\ref{y quint}), (\ref{y})) is so chosen that the  matter density parameter $\Omega_m$ and Dark Energy density parameter $\Omega_\pi$ attain respectively the values around 0.30 and 0.70 at the present epoch. It is observed that when initial value of $x$ is $\sim 0$ the results do not change significantly with the change of initial $x$, whereas the results are very sensitive to the choice of initial value of $y$. The value of $\lambda$ (Eq. (\ref{l quint}), (\ref{l})) is treated as a parameter in this work. The variable $\epsilon$ (Eq. (\ref{e})) signifies the contribution of the Slotheon term $\frac{G^{\mu\nu}}{2 M^2}\pi_{;\mu}\pi_{;\nu}$. Therefore for quintessence model we take $\epsilon=0$ and for Slotheon model a non zero value of $\epsilon$ is adopted. 

We consider an exponential form of the potential $V(X)$ as given below,
\begin{equation}
V(X)={\rm exp}\left(-\frac{C X}{M_{\rm pl}} \right)\,\,, \label{potential}
\end{equation}
where $X \equiv\phi$ or $\pi$ as the case may be. From Eq. (\ref{l quint}) or Eq. (\ref{l}) it can be seen that for the exponential form of potential, $\lambda=\frac{M_{\rm pl} V'}{V}=C$, while $C$ is a constant. A discussion on the importance of exponential form of potential to address the validity of Swampland criteria is given in Ref. \cite{patreek}. Similar form of such phenomenologically viable exponential potential for general quintessence Dark Energy has also been considered in Ref. \cite{ds} for Dark Energy equation of state analysis with CPL parametrisation in the context of discussing Swampland criteria. Therefore in what follows in this paper, general quintessence or standard quintessence refers to the general/standard quintessence model with exponential form of the potential $V(\phi)$ Eq.(\ref{potential}). The same is true for the Slotheon field Dark Energy also (exponential form of $V(\pi)$). 

\section{Generalised Parametrisation of Thawing Dark \\ Energy Model }

We investigate in this work the Swampland criteria for the case of Dark Energy from Slotheon scalar field model when confronted with the observational limits. We first obtain the observational limits for Dark Energy equation of state $\omega_{\rm DE}$ in case of two-parameter generalisation of different thawing Dark Energy models as given in Ref.   \cite{parametrisation}. The equation of state $\omega_{\rm DE}$ for such a 
two-parameter generalised thawing Dark Energy model takes the form \cite{parametrisation}
\begin{equation}
\dfrac{d \omega_{\rm DE}(a)}{da} = (1+\omega_{\rm DE}(a)) f(a)\,\,,\label{parameter0}
\end{equation}
where $f(a)=\frac{c}{a^n}$. In the above, $a$ represents the scale factor ($a=\frac{1}{1+z}$) and $c$, $n$ are the two parameters. It may be noted that for $c=1$, $n=1$ this generalised parametrisation can be reduced to CPL parametrisation form \cite{CPL}. For $n\neq1$, Eq. (\ref{parameter0}) can be reduced to the form \cite{parametrisation}
\begin{equation}
\omega_{\rm DE}(a)=-1+(1+\omega_0){\rm exp}\left[\frac{c}{n-1}(1-a^{(1-n)})\right] \,\,, \label{parameter}
\end{equation}
and it is straightforward to obtain from Eq. (\ref{parameter0}) that for $n=1$
\begin{equation}
\omega_{\rm DE}(a)=-1 +(1+\omega_0)a^c\,\,.\label{parameter2}
\end{equation}
In both the above two equations, $\omega_0$ denotes the equation of state in the present epoch. 

We consider Euclid simulations for future data acquisition \cite{euclid} and the values of $\omega_0$ in the above equations are adopted to be the different values of $\omega_0$ within 1-$\sigma$ range (Fig. 2.4  of Ref \cite{euclid}) as given in Euclid's future simulations of data. 

In the present work the region of variations of Dark Energy equation of state $\omega_{\rm DE}(z)$ with $z$ are obtained by using Eqs. (\ref{parameter}, \ref{parameter2}) within the range of values of the parameters $c$ and $n$ that obey the theoretical constraints  as given in Fig. 2 of Ref.  \cite{parametrisation}.

\section{Calculations and Results}

We first make some preliminary studies of the behaviour of different quantities related to the Slotheon Dark Energy model. To this end we show in Fig. 1 the variations of the quantity $x$ (Eq. (\ref{x})) with redshift $z$ for three different values of $\lambda$ (Eq. (\ref{l})). It can be seen from Fig. 1 that although the initial value of $x$ has been chosen close to 0, even the growth of $x$ is negligible with $x\sim0.1$ at the present epoch. This signifies that the value of $\dot{\pi}$ (Eq. (\ref{x})) remains very small. This translates to the fact that the kinetic term of the scalar field $\pi$ is negligible throughout the evolution in time. Moreover the exponential form of potential $V(\pi)$ (Eq. \ref{potential})) that we choose for this calculations, indicates that the potential $V(\pi)$ always remains positive in this case. 
\begin{figure}[H]
\begin{center}
\includegraphics[scale=0.5]{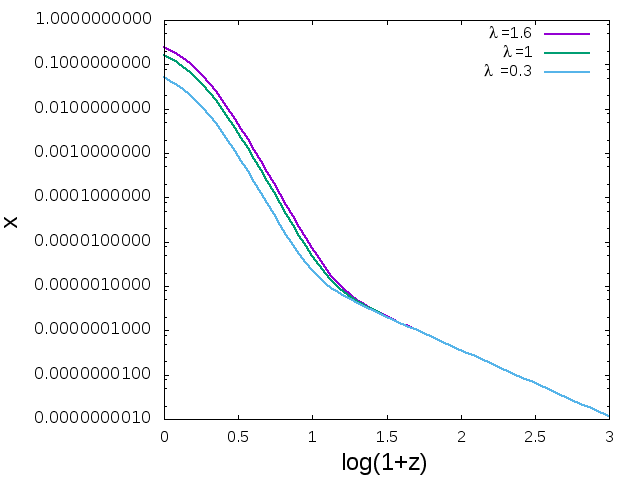}
\caption{The variations of the variable $x$ with ${\rm log}(1+z)$}
\end{center}
\end{figure}

We now calculate the variations of Dark Energy equation of state with the experimental constraints. As described in the last section, we adopt the generalised thawing Dark Energy parametrisation for obtaining the variations of Dark Energy equation of state with the parameter $\omega_0$ chosen from 1-$\sigma$ constraint as given in Euclid simulated results. We choose this our upper bound for $\omega_{\rm DE}(z)$ vs $z$ with the 1-$\sigma$ constraints for comparison with the similar variations of Dark Energy equation of states with $z$ as obtained from Slotheon model of Dark Energy with different values of $\lambda$. Similar computations are also performed for quintessence field Dark Energy equation of state and its agreement with Swampland criteria  are tested to make a comparison between Slotheon Dark Energy  model and quintessence model in terms of obeying the Swampland criteria. In this section we also compare our results of Slotheon field Dark Energy with the 1-$\sigma$, 2-$\sigma$, 3-$\sigma$ upper bounds of $\omega_{\rm DE}(z)$, constructed by CPL parametrisation with current cosmological results of Ref. \cite{9of_DE} as given in Ref \cite{l.heisenberg}.

\begin{figure}[H]
\begin{center}
\includegraphics[scale=0.5]{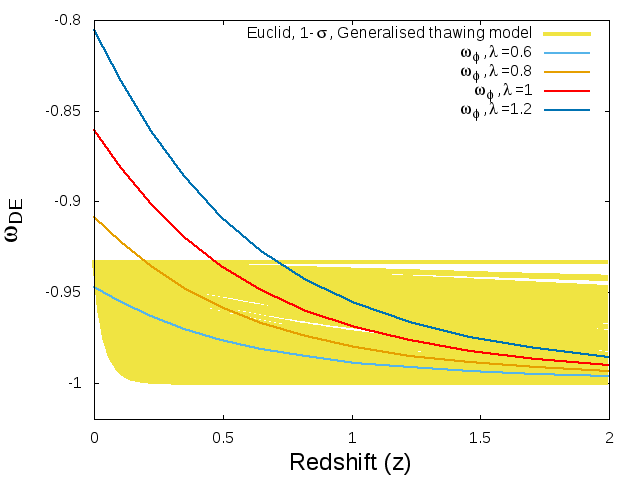} 
\caption{The variations of the upper bound of the equation of state $\omega_{\rm DE}$ with $z$ when $\omega_0$ is within the 1-$\sigma$ range of the simulated future Euclid's result and the variations of equation of state $\omega_{\rm DE}$ with $z$ for standard Dark Energy quintessence model.}\label{genquint}
\end{center}
\end{figure}

In Fig. \ref{genquint} we furnish the variations of the upper bound of the equation of state $\omega_{\rm DE}(z)$ with $z$ when $\omega_0$ is within the 1-$\sigma$ range of the simulated future 
Euclid's result and Eq. (\ref{parameter}) and (\ref{parameter2}) are adopted for the equation of state with the values of the parameter $c$, $n$ within the theoretical constraints given 
in Ref. \cite{parametrisation}. This 1-$\sigma$ range for $\omega_{\rm DE}(z)$ vs $z$ is represented by the yellow band in Fig. \ref{genquint}. The band includes all allowed 
values of $c$ and $n$ as obtained from Fig. 2 of Ref. \cite{parametrisation} with Eqs. (\ref{parameter}) and (\ref{parameter2}) where $\omega_0$ is given by 1-$\sigma$ constraints of 
Euclid results. In Fig. \ref{genquint} we also plot the calculated results of $\omega_{\rm DE}(z)$ vs $z$ 
for standard quintessence Dark Energy model for four values of $\lambda$ namely $\lambda=  0.6, 0.8, 1, 1.2$. It may be mentioned here that $\omega_{\rm DE}(z)$ is nothing but $\omega_\phi$ for quintessence scalar field as mentioned in Section 2. From Fig. \ref{genquint} it can be seen that for $\lambda=0.6$ the quintessence model lies very much within the generalised thawing Dark Energy model region for 1-$\sigma$ range of $\omega_0$ as given by the Euclid simulation of future data. It is also seen from Fig. \ref{genquint} that for $\lambda=1$, the quintessence model does not quite satisfy the yellow region indicating that the standard quintessence model does not fully obey the Swampland criteria. The situation is even worse when $\lambda>1$ for quintessence field Dark Energy. Therefore it appears from Fig. \ref{genquint} that the general quintessence model barely satisfies the Swampland criteria when compared with the thawing Dark Energy model parametrisation with $\omega_0$ is adopted to be the values within 1-$\sigma$ region as obtained from the analysis of the simulated future Euclid data.

\begin{figure}[H]
\begin{center}
\includegraphics[scale=0.5]{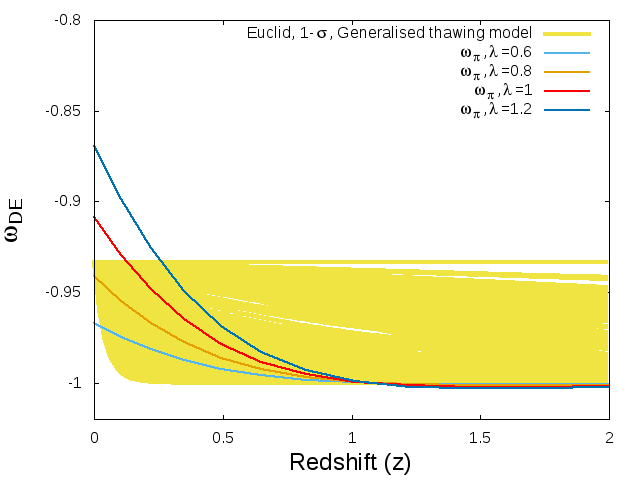}
\caption{The variations of the upper bound of the equation of state $\omega_{\rm DE}$ with $z$ when $\omega_0$ is within the 1-$\sigma$ range of the simulated future Euclid's result and the variations of equation of state $\omega_{\rm DE}$ with $z$ for the Slotheon field Dark Energy model.} \label{gensloth}
\end{center}
\end{figure}

In Fig. \ref{gensloth} we make similar comparison for Slotheon field Dark Energy model in the context that Slotheon Dark Energy obeying the string Swampland criteria. In Fig. \ref{gensloth} the yellow region is the same as that in Fig. \ref{genquint}. We calculate the equation of state $\omega_{\rm DE}(z)$ as it varies with $z$ for the case of Dark Energy from Slotheon scalar field model for four different values of $\lambda$ namely $\lambda=0.6,0.8,1,1.2$. Here, for Slotheon scalar field model of Dark Energy, the equation of state $\omega_{\rm DE}(z)$ is the same as $\omega_\pi$  mentioned in Section 2. From Fig. \ref{gensloth} it is observed that for $\lambda=0.8$ the Slotheon field model is well within the yellow region i.e., the generalised thawing Dark Energy model region with 1-$\sigma$ bound on $\omega_0$ (from Euclid). It can also be noted from Fig. \ref{gensloth} that for $\lambda=1$ the Slotheon model is marginally beyond the yellow region. From Fig. \ref{genquint} and Fig. \ref{gensloth} it is clearly observed that the Slotheon field model better satisfies the Swampland criteria than the standard quintessence model and therefore the tensions with the Swampland criteria are less severe for Slotheon Dark Energy model than the general quintessence Dark Energy model.

\begin{figure}[H]
\includegraphics[scale=0.5]{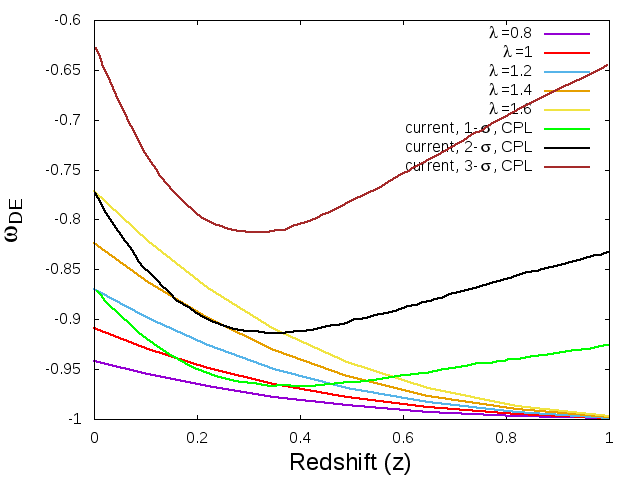}
\includegraphics[scale=0.5]{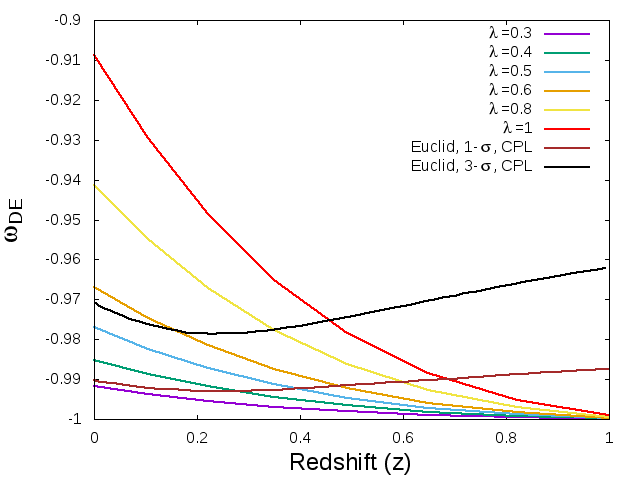}
\caption{(a) Left panel: The Swampland criteria for the Slotheon field Dark Energy is explored for different values of $\lambda$ by computing the equation of state $\omega_{\rm DE}(z)$ for
different $z$ and comparing with the experimental 1-$\sigma$, 2-$\sigma$, 3-$\sigma$ upper bounds for the $\omega_{\rm DE}(z)$ vs $z$ are adopted from Fig. 1 of Ref. \cite{l.heisenberg} where the CPL parametrisation is used with experimental constraints obtained from SNeIa, CNB, BAO and $H_0$ measurements. 
(b) Right panel: Same as the left panel but using Euclid simulated constraints with CPL parametrisation as given in Fig. 2 of Ref. \cite{l.heisenberg} } \label{gcpl}
\end{figure}

We also compare our results for Slotheon field Dark Energy and standard quintessence field Dark Energy with the 1-$\sigma$, 2-$\sigma$ and 3-$\sigma$ upper bounds on equation of state $\omega_{\rm DE}(z)$, constructed by CPL parametrisation as given in Fig. 1 of Ref.  \cite{l.heisenberg} and plot them in Fig. \ref{gcpl}(a), (left panel of Fig. \ref{gcpl}). We show the variations of $\omega_{\rm DE}(z)$ with $z$ for Slotheon scalar field model in Fig. \ref{gcpl}(a) for different values of $\lambda$, namely $\lambda=0.8,1,1.2,1.4,1.6$. It is observed from Fig \ref{gcpl}(a) that for $\lambda=1$ the equation of state for Slotheon field model lies well below the 2-$\sigma$ and 3-$\sigma$ upper bounds and therefore satisfies the Swampland criteria. It is noticed that even the equation of state for $\lambda=1.6$ is below 3-$\sigma$ upper bound. Therefore the Slotheon Dark Energy model is fully in agreement with the string Swampland criteria when the current data (from Ref. \cite{9of_DE}) as used in Ref. \cite{l.heisenberg} are considered. It can also be noted from Fig. \ref{gcpl}(a) (of this work) and Fig. 1 of Ref. \cite{l.heisenberg} that Slotheon field Dark Energy better satisfies the  string Swampland criteria than the general quintessence model of Dark Energy.

We now adopt the 1-$\sigma$ and 3-$\sigma$ upper bound of $\omega_{\rm DE}(z)$ vs $z$ plots as obtained from CPL parametrisation from Eucid simulation data as given in Fig. 2 of Ref. \cite{l.heisenberg} and compare them with the Slotheon Dark Energy equation of state for Swampland conditions. The results are furnished in Fig. \ref{gcpl}(b), (right panel of Fig. \ref{gcpl}). In Fig. \ref{gcpl}(b) we show the variations of $\omega_{\rm DE}(z)$ with $z$ for Slotheon scalar field model for different values of $\lambda$, namely $\lambda=0.3,0.4,0.5,0.6,0.8,1$. It is observed from Fig. \ref{gcpl}(b) that the plot corresponding to $\lambda=0.5$ is well below the 3-$\sigma$ upper bound of $\omega_{\rm DE}(z)$ but the situations become worse when $\lambda> 0.5$. Therefore though the Slotheon field better satisfies the Swampland criteria than the general quintessence model but it appears to be in tension with string Swampland criteria if the experimental bound is adopted to be that obtained by CPL parametrisation of $\omega_{\rm DE}(z)$ with Euclid bounds as given in Ref. \cite{euclid}. On the other hand this can easily be observed from Fig. \ref{gensloth} and Fig. \ref{gcpl}(b) that the agreement of Slotheon field with string Swampland criteria is better if generalised parametrisation of thawing Dark Energy model is adopted for the experimental bound instead of the CPL parametrisation of Dark Energy (in fact CPL parametrisation is already included within the yellow region of Fig. \ref{genquint} and Fig. \ref{gensloth}).

It may be mentioned here that to compute the plots of Fig. \ref{gensloth} and Fig. \ref{gcpl} we use the initial value of $\epsilon$ (Eq. (\ref{e})) is equal to 2.5$\times10^7$. We have also observed that for higher initial values of $\epsilon$ (say $\epsilon$= 4.5$\times10^7$, 6.5$\times10^7$ etc.), Slotheon field model better satisfies string Swampland criteria and therefore it may be concluded that by increasing the initial values of $\epsilon$, the tension of string Swampland criteria for Slotheon Dark Energy field can further be reduced.  

Although the second swampland criterion is primarily relevant and interesting in view of Dark Energy applications, we have also verified that the first of two Swampland criteria ($|\Delta \pi| < d \sim {\cal {O}}(1)$) is satisfied in the present case. For this purpose we compute $\Delta \pi$ using the relation
\begin{equation}
\Delta \pi = \sqrt{6}\int_{-\infty}^0 x dN\,\,,
\end{equation}
obtained from Eq. (\ref{x}). In our case we find that $\Delta \pi$ varies between 0.06 to 0.30 with $\lambda$ varies between 0.3 to 1.6. Therefore the first criterion is very well satisfied.

We now address the refined form of the second criterion mentioned in Section 1. It can be noted that the  criterion $\nabla^2 V/V<-c' \sim {\cal {O}}(1)$ is not satisfied by the present choice of exponential potential Eq. (\ref{potential}). But the validity of the Swampland criteria can still be addressed by exploring the validity of the condition $|\nabla_\pi V|/V > c \sim {\cal {O}}(1)$. This may be commented at the end that a refined bound on slow roll and Swampland have been proposed in Ref. \cite{chetan}. Also in Ref. \cite{9} authors discuss how the consideration of refined criterion leads to address the deviation of Dark Energy equation of state from cosmological constant.

\section{Summary and Discussions}

The string Swampland conjectures lead us to investigate how the low energy effective field theories of general relativity of gravity which appear to work well below the Planck scale can be connected to the quantum theories of gravity in a string theory landscape which are theories beyond the Planck scale. The Swampland criteria give tight constraints to the Dark Energy models of the late time acceleration of the Universe as well as on inflationary models \cite{inflation} of the early Universe. In this work we have studied the implication of string Swampland criteria on two scalar field models of Dark Energy namely general quintessence and Slotheon model, on the basis of current and future cosmological observations. The scalar field models of Dark Energy have to satisfy specially the second Swampland criterion, which is in the context of the de Sitter constraint suggests that the slope of the potential of any effective scalar field theory should be related to the potential through a constant of order one, to remain outside the Swampland. But it is observed that the general quintessence model leads to significant tension with this criterion in the view of current and future cosmological data. Therefore
in the present work we study the Slotheon Dark Energy model which is inspired by extra dimensional theories in curved space time and explore whether it satisfies the string Swampland criteria. 

In order to study the implications of string Swampland criteria on Slotheon Dark Energy model we calculate variations of Dark Energy equation of state $\omega_{\rm DE}(z)$ with redshift $z$ for Slotheon field with different values of $\lambda=\frac{M_{\rm pl} V'}{V}$. We then compare it with experimental constraints by adopting generalised parametrisation of thawing Dark Energy models and by computing variations of $\omega_{\rm DE}(z)$ with parameter $\omega_0$ where the latter is chosen from 1-$\sigma$ constraints as given in Euclid \cite{euclid} simulated future cosmological results. Similar computations are also done for standard quintessence Dark Energy model. It is noted that Slotheon field better satisfies the string Swampland criterion than quintessence field.

In addition we also compare our results (of variations of $\omega_{\rm DE}(z)$ with $z$ for Slotheon Dark Energy model) with 1-$\sigma$, 2-$\sigma$, 3-$\sigma$ upper bounds on $\omega_{\rm DE}(z)$ constructed by CPL parametrisation as given in Fig. 1 and Fig. 2 of Ref. \cite{l.heisenberg}. In Ref. \cite{l.heisenberg} current data from Ref. \cite{9of_DE} are used in Fig. 1 for the variations of $\omega_{\rm DE}(z)$ with $z$ and Euclid \cite{euclid} simulated results are also used (in Fig. 2 of Ref. \cite{l.heisenberg})
 to construct experimental upper bound of $\omega_{\rm DE}(z)$. It is clear from this work that Slotheon field Dark Energy is well in agreement with the Swampland criterion for current experimental constraints and it satisfies better the Swampland criterion than the quintessence field. It is also noted that for Euclid simulated future results, Slotheon Dark Energy model is more in agreement with the Swampland criteria for generalised parametrisation of thawing Dark Energy models than CPL parametrisation of Dark Energy. We also observe for Slotheon Dark Energy that by increasing the initial value of variable $\epsilon$ (Eq. (\ref{e})), the tension with string Swampland criteria can further be relieved.

\vspace{1cm}
\noindent{\bf Acknowledgement}

One of the authors (U.M.) acknowledges CSIR, Govt. of India for financial support through JRF scheme (NO. 09/489(0106)/2017-EMR-I).


\begin{thebibliography}{99}
\bibitem{swampland} C. Vafa,
  arXiv:hep-th/0509212.
  \bibitem{5} M.C. David Marsh,
  Phys.\ Lett.\ B {\bf 789}, 639 (2019).
\bibitem{6} S.D. Odintsov and V.K. Oikonomou,
  arXiv:1810.03575 [gr-qc].
\bibitem{8} P. Agrawal and G. Obied,
  arXiv:1811.00554 [hep-ph].
\bibitem{7}  Y.~Olguin-Trejo, S.L.~Parameswaran, G.~Tasinato and I.~Zavala,
  JCAP {\bf 1901}, no. 01, 031 (2019).
    
 \bibitem{10}E. Elizalde and M. Khurshudyan,
  arXiv:1811.03861 [astro-ph.CO].
\bibitem{11} M. Raveri, W. Hu and S. Sethi,
  Phys.\ Rev.\ D {\bf 99}, no. 8, 083518 (2019).
\bibitem{coupling}  C. van de Bruck and C.C. Thomas,
  arXiv:1904.07082 [hep-th].
 \bibitem{upala} U. Mukhopadhyay, D. Majumdar and D. Adak, arXiv:1903.08650 [gr-qc].
\bibitem{debabratada} D. Adak, A. Ali and D. Majumdar, Phys. Rev. D {\bf 88}, 024007 (2013).
\bibitem{germani}C. Germani, L. Martucci and P. Moyassari, Phys. Rev. D 
{\bf 85}, 103501 (2012).
\bibitem{criteria1}H. Ooguri and C. Vafa, Nucl. Phys. B {\bf 766}, 21, (2007).
\bibitem{criteria2}G. Obied, H. Ooguri, L. Spodyneiko and C. Vafa, arXiv:1806.08362 [hep-th].
 \bibitem{connection} H.~Ooguri, E.~Palti, G.~Shiu and C.~Vafa,
  Phys.\ Lett.\ B {\bf 788}, 180 (2019).
\bibitem{l.heisenberg}L. Heisenberg, M. Bartelmann, R. Brandenberger and A. Refregier, Phys.\ Rev.\ D {\bf 98}, no. 12, 123502 (2018).
\bibitem{thawing} R.R. Caldwell and E.V. Linder, Phys.\ Rev.\ Lett.\  {\bf 95}, 141301 (2005).
\bibitem{parametrisation} D. Adak, D. Majumdar and S. Pal, Mon.\ Not.\ Roy.\ Astron.\ Soc.\  {\bf 437}, no. 1, 831 (2014).
\bibitem{euclid}Laureijs, R. et al., ESA/SRE (2011) 12 [arXiv:1110.3193].

\bibitem{CPL}M. Chevallier and D. Polarski, Int. J. Mod. Phys. D {\bf 10}, 213
(2001); E.V. Linder, Phys. Rev. Lett. {\bf 90},
091301 (2003).
\bibitem{3} S. Vagnozzi, S. Dhawan, M. Gerbino, K. Freese, A. Goobar and O. Mena,
  Phys.\ Rev.\ D {\bf 98}, no. 8, 083501 (2018).
\bibitem{9of_DE}D.M. Scolnic et al., Astrophys. J. {\bf 859}, no. 2, 101
(2018).
 \bibitem{brahma}S. Brahma and M. W. Hossain,
  arXiv:1902.11014 [hep-th].
\bibitem{DGP}G. Dvali, G. Gabadadze, and M. Porrati, Phys. Lett. B {\bf 485}, 208 (2000).
 \bibitem{dclt1} N. Chow and J. Khoury, Phys. Rev. D {\bf 80}, 024037 (2009).
 \bibitem{dclt2} A. Ali, R.Gannouji, M. W. Hossain and M. Sami, Phys. Lett. B {\bf 718}, 5 (2012).
\bibitem{horndeski} L.~Heisenberg, M.~Bartelmann, R.~Brandenberger and A.~Refregier,
  arXiv:1902.03939 [hep-th].
\bibitem{quint tsuji} S. Tsujikawa, Class. Quant. Grav. {\bf 30}, 214003 (2013).
\bibitem{sami}E.J. Copeland, M. Sami and S. Tsujikawa, Int. J. Mod. Phys. D {\bf 15}, 1753 (2006).
\bibitem{patreek}P. Agrawal, G. Obied, P.J. Steinhardt and C. Vafa,
 Phys.\ Lett.\ B {\bf 784}, 271 (2018).
 \bibitem{ds} Y.~Akrami, R.~Kallosh, A.~Linde and V.~Vardanyan,
  Fortsch.\ Phys.\  {\bf 67}, no. 1-2, 1800075 (2019).
 
 \bibitem{chetan}S.~K.~Garg and C.~Krishnan,
  arXiv:1807.05193 [hep-th].
\bibitem{9} C.I. Chiang, J.M. Leedom and H. Murayama,
  arXiv:1811.01987 [hep-th].
\bibitem{inflation} A. Kehagias and A. Riotto,
  Fortsch.\ Phys.\  {\bf 66}, no. 10, 1800052 (2018);  J.L. Lehners,
  JCAP {\bf 1811}, no. 11, 001 (2018);  W.H. Kinney, S. Vagnozzi and L. Visinelli,
  arXiv:1808.06424 [astro-ph.CO].
  

\end{thebibliography}
\end{document}